\documentclass[fleqn,usenatbib]{rasti}

\usepackage{newtxtext,newtxmath}
\usepackage{listings}

\usepackage[T1]{fontenc}

\DeclareRobustCommand{\VAN}[3]{#2}
\let\VANthebibliography\thebibliography
\def\thebibliography{\DeclareRobustCommand{\VAN}[3]{##3}\VANthebibliography}

\usepackage{graphicx}	
\usepackage{amsmath}	

\title[vis-r]{\texttt{vis-r}: Quick radial profile modelling for radio interferometric visibilities}

\author[G. M. Kennedy]{
Grant M. Kennedy$^{1,2}$\thanks{E-mail: g.kennedy@warwick.ac.uk}
\\
$^{1}$Department of Physics, University of Warwick, Gibbet Hill Road, Coventry, CV4 7AL, UK\\
$^{2}$Centre for Exoplanets and Habitability, University of Warwick, Gibbet Hill Road, Coventry CV4 7AL, UK\\
}

\date{Accepted XXX. Received YYY; in original form ZZZ}

\pubyear{2024}

\begin{document}
\label{firstpage}
\pagerange{\pageref{firstpage}--\pageref{lastpage}}
\maketitle

\begin{abstract}
This paper describes a method for parametric radial profile modelling of radio interferometric visibility data. Image-based parametric modelling is common in the field of circumstellar debris disks, and high resolution ALMA observations make this method computationally expensive because many high resolution images must be generated and Fourier transformed. Most circumstellar disks are axisymmetric, so here a method that avoids generating images and instead models radial profiles is outlined. The two main elements are i) $u,v$ space visibility averaging, and ii) using the Discrete Hankel Transform to convert arbitrary parametric radial surface brightness profiles to visibilities. Vertical scale height can be included, but is fixed with radius; in principle this is a limitation but in practise radially-dependent vertical structure information is beyond even ALMA's capability for all but a few debris disks. Combined, these ingredients yield a method that is one to two orders of magnitude faster than image based methods and can be run on a laptop, is sufficient for most ALMA observations of debris disks, and can be used with gradient-based fitting methods such as Hamiltonian Monte-Carlo. A small software package is provided, which is dubbed \textsc{vis-r}.
\end{abstract}

\begin{keywords}
Numerical methods -- Software -- Interferometry -- Circumstellar discs
\end{keywords}



\section{Introduction}

Interferometry is a commonly used tool in astronomy. It circumvents problems that arise from the desire to have high angular resolution at a reasonable price point, but at the cost of added complexity for the end user; ``fringe visibility` is not a term that endears new users, nor is the prospect of developing methods that convert Fourier amplitudes to images, or vice versa.

The complexity of radio interferometric data, and the accessibility of facilities such as the Atacama Large Millimeter/submillimeter Array (ALMA), has driven the development of methods and software that aid the end user. For example, the \textsc{galario} package \citep{2018MNRAS.476.4527T} takes an input sky image, and given a list of antenna baseline coordinates, visibilities, and weights can return a $\chi^2$ goodness of fit metric that aids minimisation and sampling. Modelling then becomes simpler for the end user, who can focus on generating images, for example with radiative transfer packages \citep[e.g. RADMC,][]{2012ascl.soft02015D} or more analytic methods \citep[e.g.][]{2021MNRAS.504.4497C}. Indeed, in the field of circumstellar debris disks, a workflow based around image generation and Markov Chain Monte Carlo (MCMC) sampling for parametric modelling of ALMA data has become the de facto standard.

However, there are downsides to this approach. Longer  baselines and consequent greater spatial resolution requires higher resolution model images. But users typically also desire minimal flux loss, so obtain data that also has relatively short baselines.  Combined, the observational strategy means that model images must be large to contain low spatial frequency information, but have small pixels to include high spatial frequencies. Likewise the two-dimensional Fourier transforms of these images are large to cover high frequencies, and have many pixels so that low frequencies are sampled properly. For long-baseline datasets the requirement can be that the images are 8192 pixels on a side (i.e. $2^{13}$ to enable Fast Fourier Transforms). For a typical circumstellar disk most of these images can be empty because there is no structure on the largest scales, but the full images must nonetheless sit in computer memory, be Fourier transformed, and sampled/interpolated to compare with the observed data. Two additional technical issues can be that i) fitting typically does not use gradient information to improve convergence and sampling efficiency, because this would be expensive with ``black box'' image generation via radiative transfer codes, and ii) fitting is done in parallel, which may lead to significant demands on computer memory and/or require expensive many-core computing resources. While the computational resources used in fitting parametric models are not normally documented, servers with many tens of cores and hundreds of gigabytes of memory are the typical weapon of choice, and fitting runs can take sizeable fractions of a day for high resolution ALMA data. \textsc{galario} was explicitly designed to harness graphical processing units (GPUs) to help circumvent resource issues and improve efficiency, but not all users have access to research-grade GPUs. These issues could potentially be addressed by faster methods that avoid generating images, and/or more analytic methods that can implement automatic differentiation, which would enable faster and more efficient fitting.

This article outlines a method that circumvents the aforementioned problems, providing a way to model radial profiles of axisymmetric structures for high resolution interferometric datasets. There are two main ideas; one is that arbitrary radial profiles can be converted to visibilities with the Hankel Transform \citep[in this case the Discrete transform, or DHT,][]{1987CoPhC..43..181J,1994JChPh.101.3936L,2015JOSAA..32..611B}.\footnote{The reverse is also true; the the \textsc{frank} code described by \citet{2020MNRAS.tmp.1491J} uses the DHT to rapidly convert observations to radial profiles (though strictly speaking only uses the same forward transform used here).}. The second idea is that given the ability to model the visibilities directly, $u,v$ averaging of observed visibilities becomes a key expedient. A small \textsc{python} package has been written and is dubbed \textsc{vis-r}, which can be run on typical high-resolution ALMA debris disk datasets on an Apple Mac laptop in a few minutes.

\section{Methods}

This section outlines the recipe for the fast radial profile modelling of visibilities. In what follows the 860\,$\mu$m ALMA observations of HR~4796 analysed by \citet{2018MNRAS.475.4924K} are used for illustrative purposes. The data show an inclined narrow ring with a radius of 1.1\arcsec. The self-calibrated data were used, and the raw data initially averaged to four channels with time bins of 30\,s, resulting in 321,322 visibilities.

\subsection{Visibility averaging}\label{sec:avg}

The idea of binning visibilities is not new; to image interferometric data a Fast Fourier Transform (FFT) is commonly used, for which the $u,v$ data are first discretised onto a uniform grid, i.e. ``gridded''. However, for modelling the typical averaging approach is to average over time and frequency; this preserves the data structure, and can reduce the volume significantly to yield products that can be distributed for continuum modelling. For modelling via images and Fourier transforms this averaging method is sufficient as image generation is normally the most expensive step. However, for the approach outlined here to be fast, averaging in $u,v$ space (i.e. sky-projected baseline length$/\lambda$ space) is done, which removes the time and frequency dimensions completely and can reduce the visibility data volume by one to two orders of magnitude without a significant loss of information. An alternative method is baseline-dependent averaging, where data are averaged in time and/or frequency based on their baseline length rather than $u,v$ location, which achieves similar levels of compression \citep{2018MNRAS.476.2029W}.

In frequency averaging, individual $u,v$ points must not be combined too enthusiastically, which results in ``beam smearing''. This effect can be quantified in terms of flux loss for a point source some distance $\Delta \theta$ from the phase center, for which the phase shift between frequency-averaged channels must be limited such that the parameter $\beta$ is small, where
\begin{equation}
    \beta = \frac{\Delta \theta}{\theta_{\rm s}} \frac{\Delta \nu}{\nu}
\end{equation}
where $\theta_{\rm s}$ is the synthesised beam size, $\nu$ is frequency, $\Delta \nu$ is the range of frequencies being averaged. For a 1\% decrement in flux for a point source at $\Delta \theta$, $\beta \approx 0.14$ \citep{1989ASPC....6..247B}.\footnote{See the notes on bandwidth smearing on the NRAO wiki: \href{https://safe.nrao.edu/wiki/bin/view/Main/RadioTutorial}{https://safe.nrao.edu/wiki/bin/view/Main/RadioTutorial}.} Using $\theta_{\rm s} \approx \lambda/b$, this equation can be rewritten as
\begin{equation}\label{eq:avg1}
    \beta \approx \Delta \theta \left[ \frac{\Delta \nu}{\nu} \frac{b}{\lambda} \right] \, .
\end{equation}
The typical application of this equation is to take $b$ as the maximum baseline in the dataset and $\Delta \theta$ as the primary beam half-width at half-maximum (i.e. the outer ``edge'' of the data). Given the average frequency of the dataset, $\Delta \nu$ is then fixed and the data averaged to a smaller number of frequency channels.

However, this averaging is conservative; if the maximum baseline were halved (e.g. the longer baseline data discarded), then the frequency averaging could be more aggressive because $\Delta \nu$ can be doubled. This more aggressive averaging keeps the term in parentheses in equation~(\ref{eq:avg1}) constant, rather than $\Delta \nu$ constant. As $u,v = b/\lambda$, the term in parentheses can then be seen as a constant radial distance in $u,v$ space over which averaging is done, $\Delta_{u,v}$, for a given $\beta$, so we can instead write
\begin{equation}\label{eq:avg2}
    \Delta_{u,v} \approx \frac{1}{\beta \, \Delta \theta} \, .
\end{equation}

The same idea applies to time averaging, but the details differ \citep{1989ASPC....6..247B}. However, with the criterion that sky rotation does not move the offset point source by a sizeable fraction of the synthesised beam, the result is that the $\Delta \nu / \nu$ in equation (\ref{eq:avg1}) is replaced by $\Delta t / P$, where $\Delta t$ is the averaging window and $P$ the Earth's sidereal period ($\approx$24\,h). Thus, time averaging can also be seen as setting a $\Delta_{u,v}$ criterion.

One can therefore reduce the data volume by binning visibility data in $u,v$ space with a constant bin size $\Delta_{u,v}$. This is in contrast to the typical approach where data are time and frequency averaged, which effectively has a bin size that decreases with $u,v$. Because the density of baselines in $u,v$ space is higher for smaller $u,v$, the former approach can reduce the number of visibilities being fit during modelling drastically, without any loss of fidelity. While this reduction may have little effect for image-based modelling, where generating images is commonly the most expensive step, it is a key expedient when modelling the visibilities directly.

To be clear, the above ``derivation'' is heuristic, not quantitative. Users should experiment for their particular case, verifying that the results do not change as $\Delta_{u,v}$ is decreased. An example case is given below, for which $\Delta \theta \gtrsim 2r$, where $r$ is the radius of an annular disk, was found to be sufficient.

\begin{figure}
    \centering
    \hspace{-0.5cm}\includegraphics[width=0.5\textwidth]{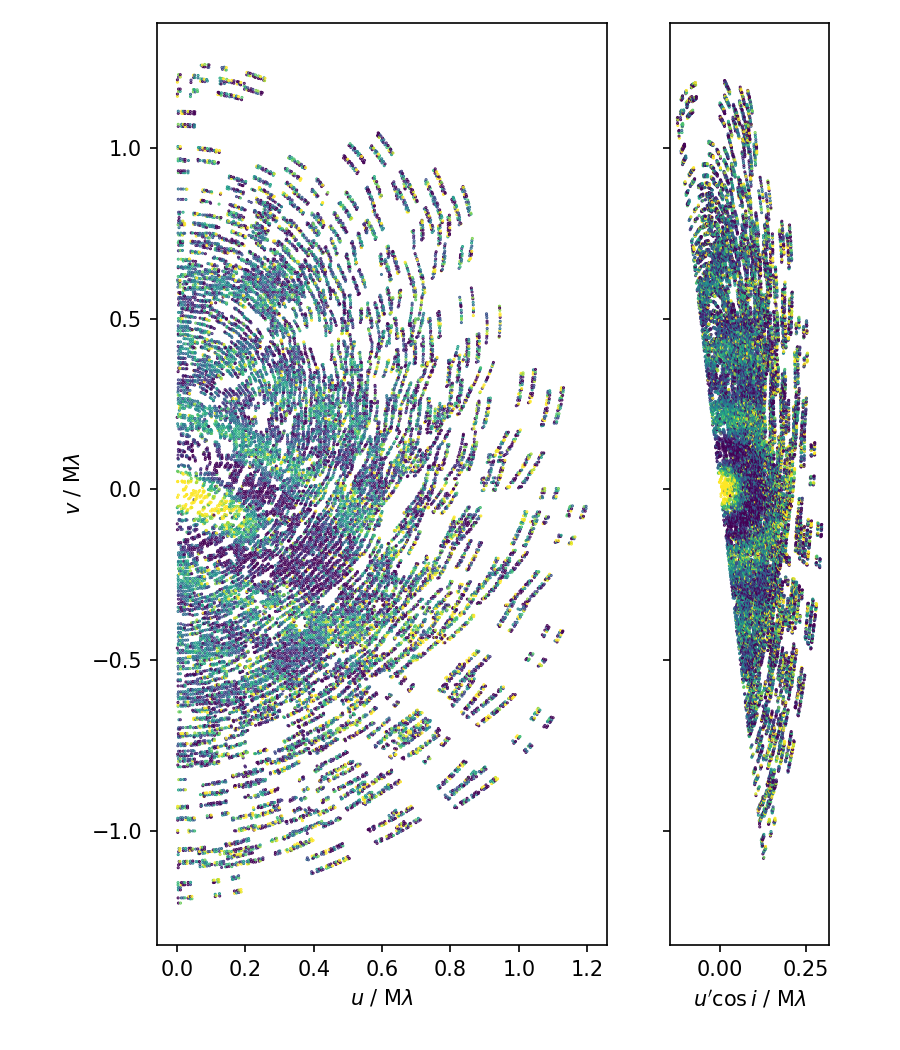}
    \caption{Visibility binning and transformation for the HR~4796 data, each point is a baseline that is coloured by the real component of the visibility in the range -0.005 to 0.01\,Jy. The left panel shows the binned visibilities ($u, v$) with $\Delta_{u,v}=7350$, and the right panel shows them after rotation and shrinking given the disk position angle and inclination ($u' \cos i, v'$). The coordinates are compressed by a factor 0.24 along the rotated $u$ axis because the disk is highly inclined. The visibility pattern is circular in the right panel, so can be modelled purely in terms of the radial dependence about the origin.}
    \label{fig:bin}
\end{figure}

In practise this binning is a weighted sum of the baseline lengths $u$ and $v$ and visibilities $V$ that fall in a given bin (i.e. $\bar{x} = \Sigma x_j w_j / \Sigma w_j$). It is not strictly the same as the ``cell-summing'' gridding method \citep[e.g.][]{1974AJ.....79...11T}, since here the resulting $u,v$ grid is not exactly regular because the $u,v$ locations in each bin are also the weighted averages. The weights $w$ in a bin are themselves summed, as they are the inverse variance for a given visibility. Baselines with negative $u$ can be rotated by 180$^\circ$, and their visibilities complex conjugated before binning, which further reduces the number of binned visibilities. An illustration of the binned visibility coordinates is shown in the left panel of Figure \ref{fig:bin} for $\Delta_{u,v}=7350$ ($\Delta \theta = 4$\arcsec). The number of visibilities in this case has been reduced from 321,322 to 19,493, and the impact of this reduction on computational cost is discussed below.

\subsection{Radial profile modelling}

This section outlines how to model a set of visibilities for a given input radial profile. Here the model will be called a ``disk'', primarily because this model has been developed for modelling of debris disks observed with ALMA, but could be any asisymmetric structure. The radial profile will necessarily be parameterised with the goal of deriving best-fit values and uncertainties for those parameters. The model could in principle be the surface brightness of a series of radial locations, but for extracting empirical radial profiles other methods are far better suited, most notably \textsc{frank} \citep{2020MNRAS.tmp.1491J}. \textsc{frank} uses the Discrete Hankel Transform to quickly convert a set of visibilities into a radial profile; the method here is essentially the reverse, converting a profile radial profile to visibilities with the DHT for comparison with the data.

To be clear, modelling radial profiles with the Hankel Transform is not new \citep[e.g.][]{2014A&A...563A.136M,2016ApJ...818L..16Z,2018ApJ...869L..48G}, though use of the DHT is to this author's knowledge novel for forward modelling of radial profiles. The DHT reduces the Hankel Transform integral to a matrix-vector multiplication, which is simple to understand and implement. Previous methods appear likely to have computed the integral numerically, perhaps using Fast Fourier Transform (FFT) based methods \citep[e.g.][]{2000MNRAS.312..257H}, which are not necessarily slower than the method discussed here. As discussed below the DHT is not actually the bottleneck here, so the specific implementation of the Hankel Transform is not critical for speed.

\subsection{Discrete Hankel Transform}

The basis for the DHT is the Hankel Transform, which is a specific case of the 2-dimensional Fourier Transform when the spatial (or frequency domain) signal is azimuthally symmetric, i.e. can be described by a radial profile. Following the notation of \citet{2020MNRAS.tmp.1491J}, the transform from spatial ($r$) to frequency ($q$) coordinates is
\begin{equation}\label{eq:hankel}
    V_\nu(q) = \int I_\nu(r) J_0(2\pi q r) 2 \pi r \, dr.
\end{equation}
Under the assumptions that the spatial signal is zero beyond some radius $R_{\rm out}$, and that the visibilities are zero beyond $Q_{\rm max}$, one can write $I_\nu$ as a Fourier-Bessel series to obtain
\begin{equation}\label{eq:dhtsum}
    V_\nu(q) = \sum_{k=1}^\infty \frac{1}{\pi Q_{\rm max}^2 J_1^2(j_{0k})} J_0 \left( \frac{j_{0k} q}{Q_{\rm max}} \right) I_\nu \left( \frac{j_{0k}}{2 \pi Q_{\rm max}} \right) \, .
\end{equation}
By truncating the series at $k=N$, the visibility can be computed at arbitrary $q$ from the intensity at a series of specified radial locations $r_k=j_{0k}/(2\pi Q_{\rm max})$, where $j_{0k}$ is the $k$th root of $J_0(r)$.

\subsection{Visibility Model}

The DHT then can be expressed as a matrix-vector multiplication,
\begin{equation}\label{eq:dht}
    V_\nu(q) = \boldsymbol{Y} \, I_\nu(r_{\rm k}, p) \, ,
\end{equation}
where the matrix $\boldsymbol{Y}$ is an $N_{\rm vis} \times N$ matrix made from the first two parts in the sum of equation \ref{eq:dhtsum}, and $I_\nu$ is a length $N$ vector of radial surface brightness at points $r_k$, the specific form of which depends on the model prescription (e.g. a Gaussian) and some model parameters $p$.Typically $p$ will contain geometric parameters that are common to every model (position angle $\phi$, inclination $i$, and sky offsets $\Delta \alpha$, $\Delta \delta$), and disk-specific parameters that dictate the surface brightness or total flux, shape, and scale of the profile (e.g. Gaussian location $r_0$ and width $\sigma_{\rm r}$). 

While the model normalisation could be done with a parameter that specifies the surface brightness of $f$ (e.g. at $r_0$), in practise it is better to use this parameter to specify the total flux at zero baseline $V_\nu(0)$. This quantity will vary little from model to model and is only weakly correlated with other model parameters.

For disk modelling the geometry is typically parameterised as a position angle $\phi$ (East of North) and inclination $i$. To account for an inclined disk the model $u,v$ distances could be radially stretched along the disk minor axis, to account for the smaller spatial scale (greater spatial frequencies). In practise however, the model is purely radial so the observational $u,v$ distances are instead squeezed. This transform is therefore achieved with a rotation and scaling of the observed visibility coordinates $u,v$ \citep[e.g. as in][]{2018MNRAS.476.4527T}, given by
\begin{gather}
    u' = u \cos \phi - v \sin \phi \notag \\
    v' = u \sin \phi + v \cos \phi
\end{gather}
and 
\begin{equation}\label{eq:trans}
    q_{\rm u,v} = \sqrt{[u' \cos i]^2 + v'^2} \, .
\end{equation}
The result of this transformation, before the hypotenuse $q_{\rm u,v}$ is calculated, is shown in the right panel of Figure \ref{fig:bin}.

\begin{figure}
    \centering
    \hspace{-0.5cm}\includegraphics[width=0.5\textwidth]{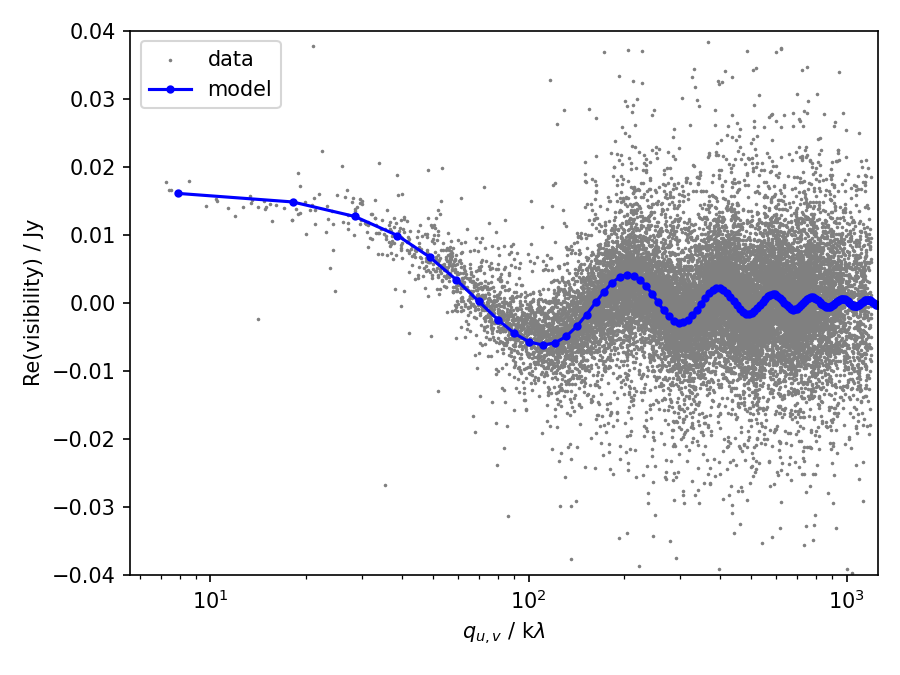}
    \caption{Model ($V_\nu(q)$) and binned data in radial visibility coordinates.}
    \label{fig:model}
\end{figure}

Given these transformed radial $u,v$ coordinates, the model is interpolated to obtain model visibilities for every observed baseline
\begin{equation}
    V_\nu(q) \rightarrow V_\nu(q_{\rm u,v})
\end{equation}
(variables other than $q$ or $q_{u,v}$ that set $V_\nu$ are omitted for clarity). A comparison of a radial model with the observed visibilities is shown in Figure \ref{fig:model}. The model is a narrow annulus with a radius of about 1.1\arcsec.

Practically, two parameters must be specified in setting up the DHT, $Q_{\rm max}$ and $N$, and the length $N_{\rm vis}$ vector of model visibilities $q$ must also be chosen. Following \citet{2020MNRAS.tmp.1491J}, $N=300$ is used, and from the data $Q_{\rm max} = \max(\sqrt{u^2+v^2})$. Then $N_{\rm vis} = \lceil (Q_{\rm max}+\Delta_{u,v}/2)/\Delta_{u,v} \rceil + 1$ points in $q$ are chosen\footnote{The function $\lceil x \rceil$ is \texttt{ceil}$(x)$, rounding the real number $x$ up to the nearest integer.}; the first $q$ is zero, and the rest span $\Delta_{u,v}/2$ to $\approx$$Q_{\rm max}$ in steps of $\Delta_{u,v}$.


The penultimate step is to incorporate vertical structure. For optically thin disks this step is absolutely necessary, as inclined disks are brightened at their ansae by the longer line of sight through those parts of the disk. In image space, a constant absolute scale height for a Gaussian distribution can be implemented with a 1-d convolution along the disk minor axis. The dispersion of the Gaussian is $\sin i$ times the real vertical dispersion to account for the disk inclination (i.e. a face-on disk has no ansae brightening so needs no convolution). The properties of Fourier transforms mean that this Gaussian convolution is simply an exponential multiplication of the ``flat'' interpolated model in Fourier space;
\begin{equation}\label{eq:vert}
    V_\nu(q_{\rm u,v})' = V_\nu(q_{\rm u,v}) \, \exp{(-[2 \pi (z_{\rm h} r_0 \sin i) \, u']^2/2)}
\end{equation}
where $z_{\rm h}$ is the relative scale height at some representative disk radius $r_0$. Other functions describing the vertical structure could also be used within this framework, as long as they are independent of radius, but the Gaussian is preferred as both astrophysically realistic \citep[e.g.][]{2019AJ....157..135M} and because it has an analytic Fourier Transform.

Typically it is assumed that the vertical density distribution in a disk is Gaussian, and that the height varies as a function of stellocentric radius, for example if dust particles have a fixed inclination. Here however the scale height $H$ is constant simply because it can be modelled analytically. In practise our ability to quantify the scale height dependence is limited, so this choice is not a serious limitation. For example, \citet{2023MNRAS.524.1229T} find that for 49~Ceti, while a constant relative scale height is preferred (i.e. $H/r=$ constant), a constant absolute height is consistent with the ALMA data at the 2$\sigma$ level. As noted by \citet{2023MNRAS.524.1229T}, this disk is one of the most promising for quantifying the radial dependence of scale height. While a radially-dependent scale height could be incorporated using the method of \citet{2023MNRAS.524.1229T}, the constant absolute vertical profile is much simpler and should be sufficient for all but the highest resolution data for the broadest disks.

Finally, the model can be translated to account for any offset of the disk from the observation phase center. This is simply a phase shift (rotation into the complex plane) of the model visibilities using the original binned $u,v$ coordinates $\gamma = u \Delta \alpha + v \Delta \delta$.
\begin{equation}\label{eq:phase}
    V_\nu(q_{\rm u,v})'' = V_\nu(q_{\rm u,v})' \, \exp{(2 \pi \iota \gamma)}
\end{equation}
where here $\iota = \sqrt{-1}$.

Some or all of the above steps used to create a single component can be used to generate more complex structures, e.g. multiple rings, each of which could have independent scale heights and sky geometry. Similarly, multiple observations of the same source can be modelled together, but allow for different relative astrometric offsets between datasets.

In summary, the procedure to compare a model with parameters $p$ with data is to choose $Q_{\rm max}$ and $N$, obtain the radial profile at $r_{\rm k}$, choose $q$ and apply the DHT (equation [\ref{eq:dht}]) to get $V_\nu(q)$. The binned observational baseline coordinates are transformed according to the disk geometry to find $q_{u,v}$, and then used to interpolate $V_\nu(q)$ at $q_{u,v}$, giving $V_\nu(q_{u,v})$. This model is then multiplied by the exponential in equation (\ref{eq:vert}) to add the vertical dimension. The model is shifted some distance from the phase center to obtain the final model $V_\nu(q_{u,v})''$. This model can then be compared with the data, e.g. by computing a sum of squared differences. To visualise how well the model fits the data, the model can be interpolated at the original unbinned $u,v$ locations and subtracted from the data, which can then be imaged.

There are two main limitations of this method. As noted above, the inability to model a radially dependent scale height is one. The other is that the loss of sensitivity towards the edges of the antenna primary beam, which is simply a Gaussian multiplication for image modelling, cannot be included easily here. Care should therefore be taken when interpreting fitting results for disks that are a sizeable fraction of the primary beam (which is 19\arcsec~FWHM at 1\,mm for ALMA's 12\,m antennas), where for example the model flux may be underestimated (see below), or surface brightness power-law indices might not be as steep as suggested.

\subsection{ Implementation and Verification}

As proof of concept some results are given for implementations in \textsc{python} with the MCMC code \textsc{emcee} \citep{2013PASP..125..306F}, and the (gradient based) Hamiltonian Monte Carlo code \textsc{stan} \citep{2017JSS....76....1C}, with the goals of showing that the results are equivalent to those obtained with an imaging code \citep[e.g.][]{2021MNRAS.504.4497C}, illustrating the effect of visibility averaging, and discussing execution times for a typical use case. The code developed for these purposes is available on GitHub.\footnote{\href{https://github.com/drgmk/vis-r}{https://github.com/drgmk/vis-r}.}

In what follows the HR~4796 ALMA dataset is modelled with a single 3-dimensional Gaussian torus that has parameters $\Delta \alpha$, $\Delta \delta$ (sky offsets), $\phi$, $i$ (position angle and inclination), $F$ (flux), $r_0$, $\sigma_{\rm r}$, and $\sigma_{\rm h}$ (disk radius, width and scale height). These parameters are the same between fitting methods, with the exception that $F$ is total flux for image modelling and the flux on the shortest baseline for visibility modelling, and $h$ is relative $H/r$ for image modelling and absolute $H/r_0$ for visibility modelling. The visibility data are averaged as above, with $\Delta_{u,v} = 7350$ ($\Delta \theta = 4$). The dataset here has a 0.17\arcsec~beam, which is at the higher end of what is possible with ALMA for debris disks, which are typically limited by surface brightness.

In all cases the Monte Carlo chains are run for 1400 steps, with the final 400 steps used to build posterior distributions for each parameter. For imaging and the \textsc{emcee} implementation of \textsc{vis-r}, 36 walkers are used, while for the \textsc{stan} implementation 6 chains are used.

\subsubsection{Consistency checks}

\begin{figure}
    \centering
    \includegraphics[width=0.5\textwidth]{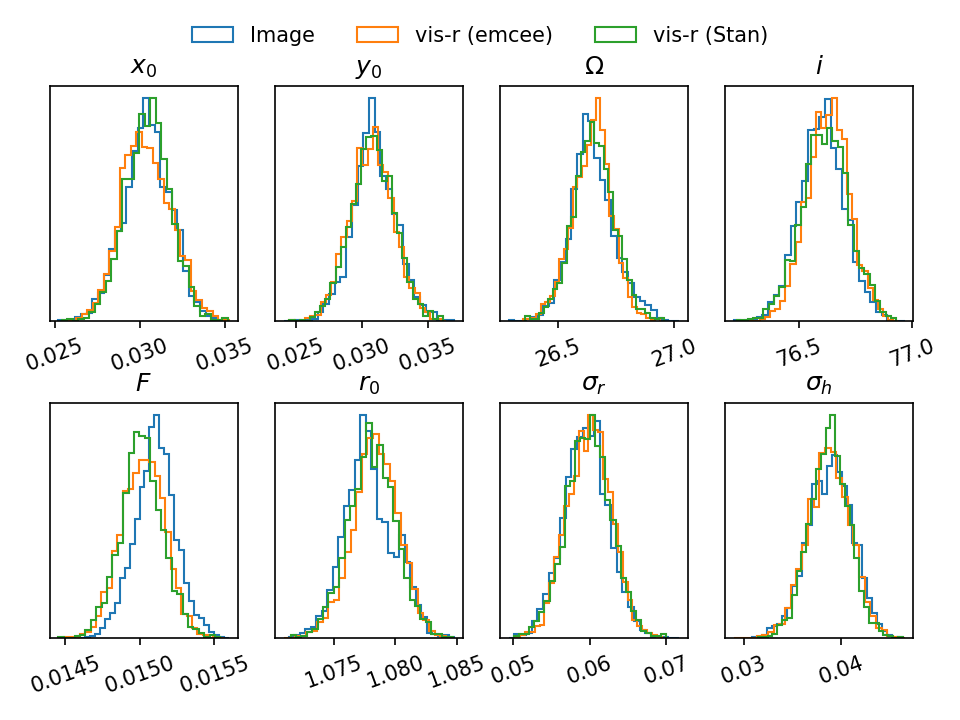}
    \caption{Comparison among image and visibility based posterior distributions for the HR~4796 dataset. Units are seconds of arc, Jansky, and degrees. The only significant difference is that the visibility method yields a lower average $F$ parameter, because for this method $F$ is the flux on the shortest baseline, while for the imaging method $F$ is the total disk flux.}
    \label{fig:comp}
\end{figure}

The first comparison is among the posterior distributions for sampling runs for the three implementations. Figure \ref{fig:comp} shows posterior distributions of the eight disk model parameters for the three methods. The only clear difference is that for the $F$ parameter the visibility model fluxes are about one percent lower, which is a result of the ALMA primary beam not being incorporated in \textsc{vis-r}, and the primary beam power being about 99\% at the disk ansae. This difference is small compared to the typical absolute flux calibration uncertainty of about 10\%. Otherwise, all parameter medians are within 1\% of the values from the image-based fitting, and the posterior standard deviations within 10\%. Thus, the visibility model is capable of producing equivalent results without generating images. Cases where the radial scale height dependence is strongly constrained in the data will require an image-based method, but as noted above these will be rare.

Similar tests for injected models are also made, using an empty visibility dataset created by subtracting the best-fit model for HR~4796. Models with randomly chosen parameters are then generated. For one test, the visibilities generated by the \textsc{stan} and \textsc{emcee} versions are compared and found to be consistent. For another test, a random model is injected into the empty dataset, and fit with the \textsc{emcee} version, with the best fit parameters found to be within 1\% of those injected. In a final test, a random model is injected into the empty data and fit with both the \textsc{stan} and \textsc{emcee} versions, and the best fit parameters found to agree within 1\%. Each test is run successfully ten times; these tests are included in the \textsc{vis-r} package and run via the \textsc{pytest} module.

\subsubsection{Effect of Averaging}

\begin{figure}
    \centering
    \includegraphics[width=0.5\textwidth]{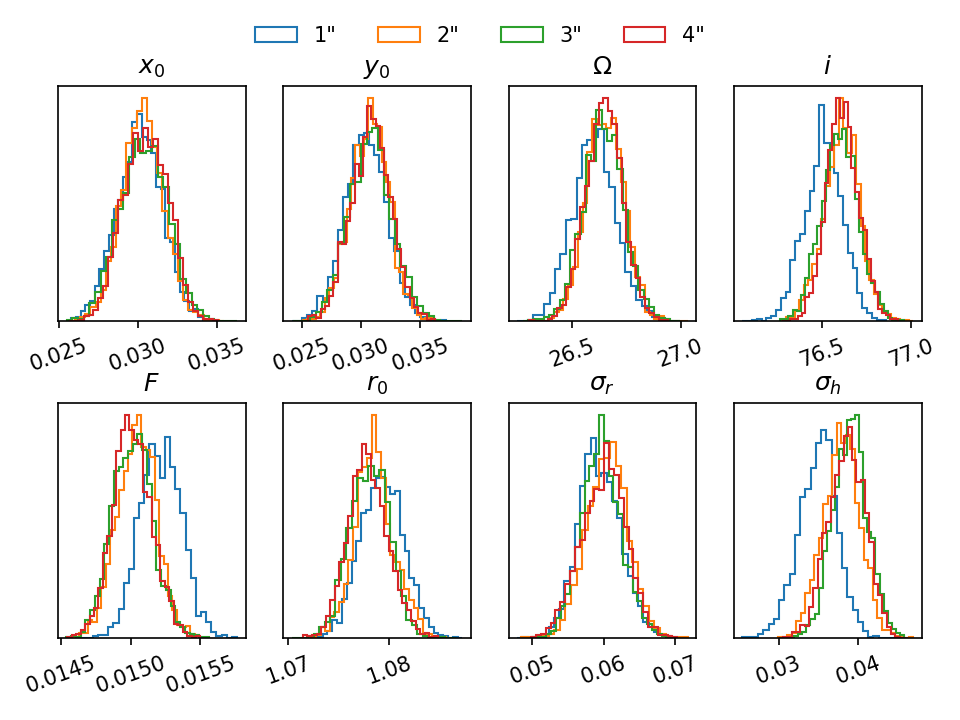}
    \caption{Effect of visibility averaging levels, with values for $\Delta \theta$ given in the legend at the top; the disk here is 1\arcsec~in radius. Averaging on the scale of the disk changes the model flux, but only by a few percent. Units are seconds of arc, Jansky, and degrees.}
    \label{fig:avg}
\end{figure}

A second comparison is among different levels of averaging, as shown in Figure \ref{fig:avg}. The averaging here has been cast in terms of seconds of arc, using equation \ref{eq:avg2} with $\beta=0.14$. It is clear from the posteriors that averaging to a spatial scale that is the same radius as the disk is somewhat detrimental, but only the flux is strongly affected, and only at the few percent level. While one expects the flux to decrease with harder averaging, the model flux actually increases in this case; the flux on the shortest baseline does indeed decrease, but the flux here is that for zero baseline, for which no data exist, so the increase seems to be an artefact of how the averaging affects the visibility curve as a function of $q_{u,v}$. Reliable initial parameter estimates can therefore be quickly obtained with fairly hard averaging. This is indeed the suggested approach; a quick assessment of models with hard averaging ($\Delta \theta \sim r_{\rm disk}$), followed by fitting with less averaging to obtain final parameters.

\subsubsection{Speed}

\begin{figure}
    \centering
    \includegraphics[width=0.45\textwidth]{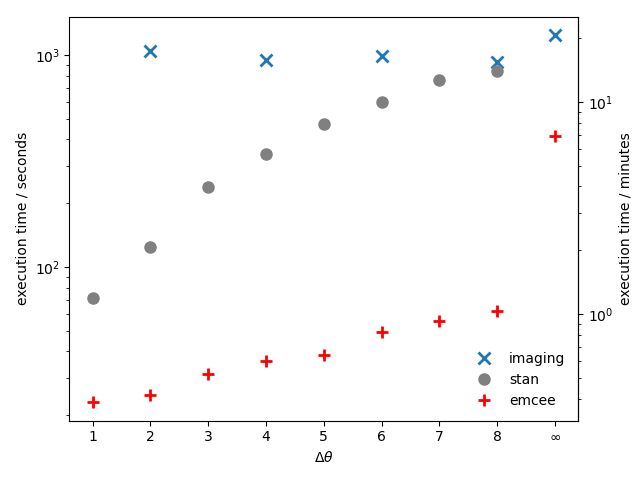}
    \caption{Execution times for imaging, and the \textsc{emcee} and \textsc{stan} implementations of \textsc{vis-r}. The x-axis, $\Delta \theta$, dictates how hard the visibility averaging is, and therefore how fast \textsc{vis-r} runs. $\Delta \theta = \infty$ corresponds to no averaging. The time for the \textsc{stan} implementation in this case is not shown, but is about 5.5\,h.}
    \label{fig:timing}
\end{figure}

To give a feel for the speed of these methods, the time taken for one model likelihood evaluation with this dataset on an Apple M2 MacBook Air in \textsc{python} is about 50\,ms for the imaging code. The \textsc{vis-r} method in \textsc{python} takes about 1\,ms, so is significantly faster. With no binning of visibilities this time increases to about 25\,ms, illustrating why binning is beneficial. In \textsc{stan} a single evaluation with \textsc{vis-r} takes about 15\,ms, but bear in mind that this time includes calculation of the gradients ($d \chi^2 / dp_i$), and that convergence is much faster and fewer samples are needed to build posteriors. For \textsc{vis-r} the most expensive steps are transformation of the visibility coordinates and interpolation (i.e. the DHT is not actually the bottleneck).

To give some numbers for fitting, Figure \ref{fig:timing} shows the time taken for full fitting runs with the three codes. These are not strictly equivalent comparisons, as the \textsc{stan} implementation first estimates the parameter posteriors using the Pathfinder algorithm \citep{JMLR:v23:21-0889}\footnote{HMC requires a ``metric'', which can be the inverse of the covariance matrix that describes parameter scales and correlations. Most of the ``warmup'' time spent by \textsc{stan} is learning this metric, starting with the assumption of a unit diagonal, so sampling is significantly faster if all parameters are initially scaled to have unit variance. The posterior estimates from Pathfinder, which is a Variational Inference algorithm, are used to find these initial scalings. The Pathfinder runs are a small fraction of the total execution time, highlighting that posterior estimates with Variational Inference are also a viable and rapid path to results.}, but this step aids sampling significantly so it is expected that this step would normally be included. The timing for the image based fitting is approximately constant, because most of the time is spent generating images. The \textsc{emcee} fitting is very fast, so fast that it appears that there are fixed overheads that limit the code from running faster for small $\Delta \theta$; the number of visibilities scales as $\Delta \theta^2$, but the timings do not scale this way, suggesting that with optimisation the code may run even faster. The \textsc{stan} implementation is several times slower than \textsc{emcee}, but this is expected because \textsc{stan} is also computing likelihood gradients. For \textsc{stan} the scaling with $\Delta \theta$ is closer to that expected.


The comparison in Figure \ref{fig:timing} is fair in one sense; if one wants to quickly estimate some model parameters for a handful of models, subtract those models from the data, and assess which model is sufficient to explain the data, then the \textsc{emcee} implementation is clearly superior. However, if one wants to obtain robust posterior distributions with confidence that the Monte Carlo chains have converged and that there are a sufficient number of independent samples, then the \textsc{stan} implementation is much better than Figure \ref{fig:timing} suggests. The reason is that the samples from \textsc{emcee} are correlated, whereas those from \textsc{stan} (i.e. HMC) are nearly independent. To give a specific example, for the examples above, the \textsc{emcee} chains have an estimated autocorrelation time of 100 steps, meaning that if the final 400 steps were taken to build posteriors, there would be approximately $400*36/100=144$ independent samples. For HMC the effective sample size is about the number of samples, so there are 2400 samples. Thus, in this case the \textsc{emcee} implementation should be run for 17 times more steps than the \textsc{stan} version for a fair comparison in terms of effective sample size, in which case the two \textsc{vis-r} codes are similarly fast. For the example in Figure \ref{fig:timing}, the \textsc{stan} implementation is actually faster for smaller $\Delta \theta$ if effective sample size is the metric chosen.

A final note on speed comparisons is that with gradient based methods one need not do a full HMC sampling run; finding a best fit model to subtract from some data does not require uncertainties on those best fit parameters, and converging to a best fit model can be very fast with gradient information. Variational Inference can provide useful posteriors more rapidly than HMC, useful in two ways here. One is that posterior estimates can be made in about the same time as those with \textsc{emcee} in Figure \ref{fig:timing}, enabling fast model exploration. Another is for cases where \textsc{vis-r} with MCMC or HMC might become prohibitively slow, e.g. due to a large number of model parameters.

\section{Examples}

Two example fitting results are shown here; those for HR~4796 as modelled above, and the protoplanetary disk AS~209 as observed by the Disk Structures at High Angular Resolution Programme (DSHARP) \citep{2018ApJ...869L..48G}. For HR~4796 the residuals are consistent with noise, and appear essentially identical to the residuals from the image-based modelling presented in \citet{2018MNRAS.475.4924K} (their Figure 3).

\begin{figure}
    \centering
    \hspace{-0.5cm}\includegraphics[width=0.5\textwidth]{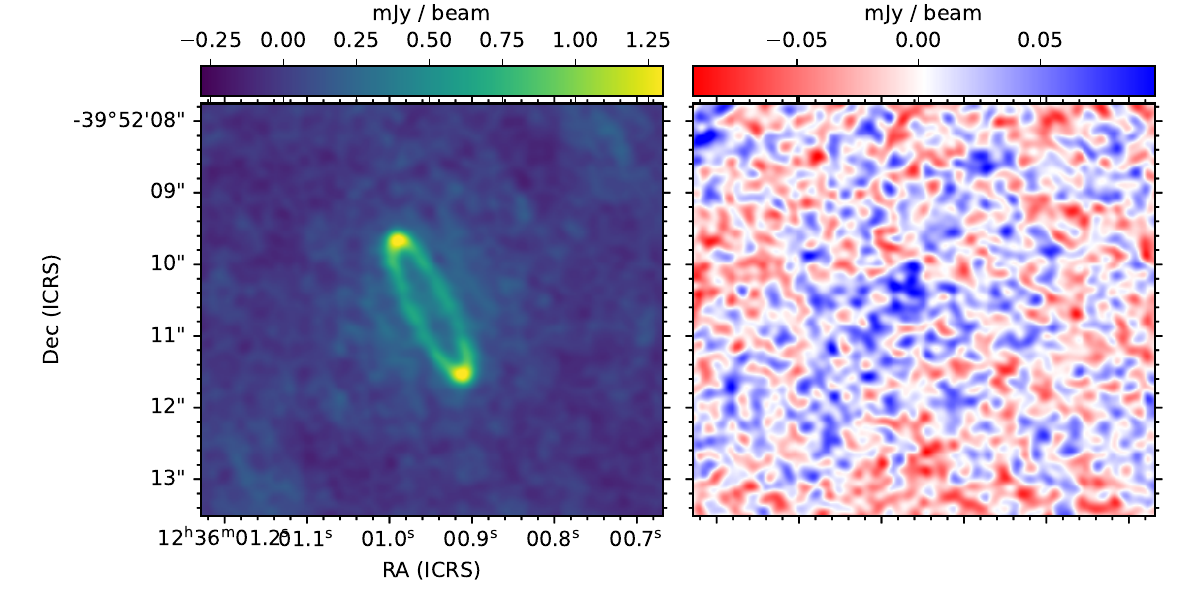}
    \caption{Data (left panel) and residuals (right panel) for the \textsc{vis-r} model of HR~4796.}
    \label{fig:hr4796resid}
\end{figure}

For AS~209 the model has seven concentric Gaussian rings that share the same center and geometry (21 parameters), a central Gaussian component (2 parameters), and offset and disk geometry (4 parameters), giving 27 parameters in total. The residuals are to be compared to Figure 4 of \citet{2018ApJ...869L..48G}, and there are again only minor differences. The differences in common model parameters are generally very small, with the only notable difference being that the outer ring (B140) has radius 132\,au and width 14\,au, rather than radius 139.1\,au and width 23.1\,au. This ring is however faint and blended with a brighter ring at 120\,au, so unlikely to be well constrained (or actually Gaussian in shape). Other rings are within an au or so in terms of radius and width.

The AS~209 data have $2 \times 10^6$ visibilities after the data are averaged to 8 channels for each of the four spectral windows. With $\Delta \theta = 2$\arcsec (the outermost ring is at 1.15\arcsec) there are $10^5$ binned visibilities. With 27 parameters the fitting is slower than for simpler cases because i) with many rings a single model evaluation is slower, ii) with 27 parameters \textsc{emcee} requires at least 54 parallel MCMC chains ("walkers") iii) convergence is slower, and most importantly iv) finding initial parameters that yield sensibly converged models is time consuming, taking longer than the final fitting to estimate posteriors. Final fitting runs of 20,000 MCMC steps with 54 walkers took about an hour on an M2 MacBook Air.

\begin{figure}
    \centering
    \hspace{-0.5cm}\includegraphics[width=0.5\textwidth]{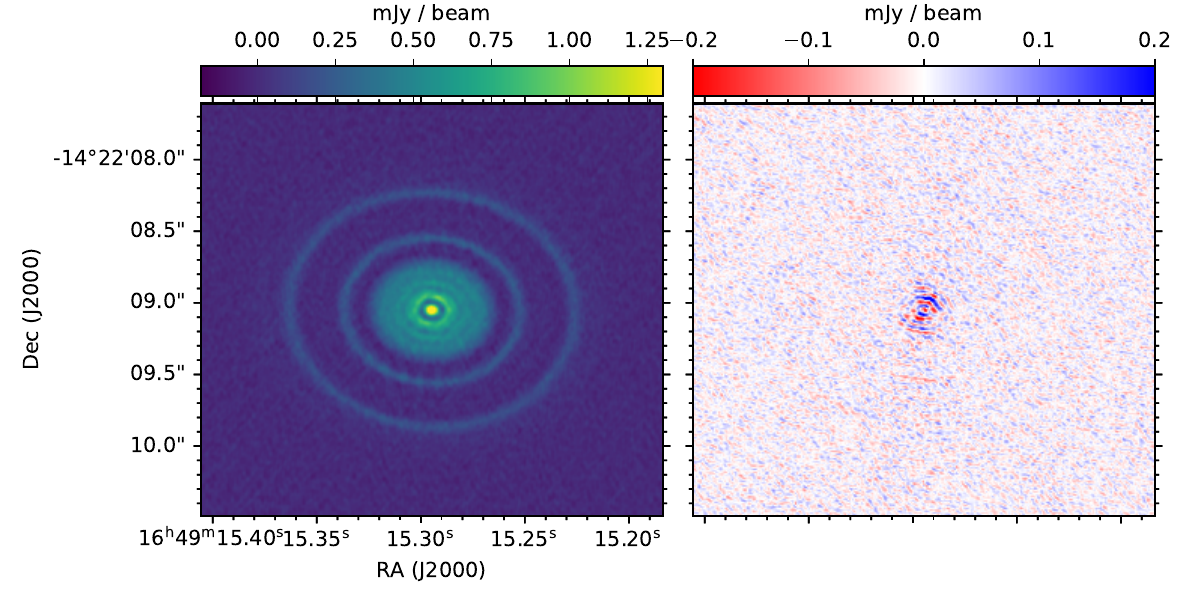}
    \caption{Data (left panel) and residuals (right panel) for the \textsc{vis-r} model of AS~209.}
    \label{fig:as209resid}
\end{figure}

\section{Summary and Outlook}

This paper outlines a method for the rapid radial profile modelling of interferometric data, dubbed \textsc{vis-r}. The main elements are visibility averaging in $u,v$ space and the Discrete Hankel Transform to convert model profiles into visibilities. Aside from cases where vertical structure is key, or disks that are large compared to the primary beam, \textsc{vis-r} quickly yields results that are equivalent to much more complex and computationally expensive image-based methods. Others have used similar methods in the past, but the inner workings are not normally described in detail, nor the code made public. Some code that implements \textsc{vis-r} is available on GitHub, consisting primarily of a \textsc{python} script that can model a dataset with \textsc{emcee} or \textsc{stan} implementations given a choice of radial profile and some input parameters.

An important benefit is that the method can be implemented for use with Hamiltonian Monte Carlo and other gradient-dependent methods, such as Variational Inference; the benefits are that posterior estimates can be found very rapidly, and posteriors can be generated from sampling runs whose samples are essentially independent. While the \textsc{emcee} version yields apparently satisfactory results much more rapidly, the non-independence of subsequent samples with MCMC means that the two implementations actually have similar run times for a fixed effective number of samples. Thus, a suggested approach is to use the \textsc{emcee} version (including heavy $u,v$ averaging where necessary) to find good parameter estimates with relatively short MCMC runs (a few thousands of steps), and then obtain final posteriors with a similar number of steps with the \textsc{stan} implementation, or many more MCMC steps with the \textsc{emcee} version.

The method described here was inspired by the modelling requirements for the ongoing ALMA Large Programme ARKS (Marino et al. in prep), so has naturally been used on debris disk observations. The disk structures are relatively simple, so models do not require too many radial components and are fast to run on any dataset. Thus, relatively little effort was put into optimisation; working in visibility space combined with averaging naturally makes for fast modelling. Protoplanetary disks can be significantly more complicated, which is reflected in the effort needed to model them, as shown by the AS~209 example above. Thus, there may be merit in further development for more complex modelling efforts. This effort could be technical, e.g. no attempt has been made here to use the GPU that is now in many Mac laptops. Methodological development might include incorporating the primary beam or radially-dependent vertical structure. Both use, and development in any direction, are of course encouraged.

\section*{Acknowledgements}

This work benefited from thoughtful comments and encouragement from the referees. The method outlined here is inspired by \textsc{frank} \citep{2020MNRAS.tmp.1491J} and data from the ALMA Large Programme ARKS. GMK was supported by the Royal Society as a Royal Society University Research fellow.

\section*{Data Availability}

The code developed for this paper is available on GitHub at \href{https://github.com/drgmk/vis-r}{https://github.com/drgmk/vis-r}. This paper makes use of the following ALMA data: ADS/JAO.ALMA\#2016.1.00484.L, ADS/JAO.ALMA\#2015.1.00032.S. ALMA is a partnership of ESO (representing its member states), NSF (USA) and NINS (Japan), together with NRC (Canada), MOST and ASIAA (Taiwan), and KASI (Republic of Korea), in cooperation with the Republic of Chile. The Joint ALMA Observatory is operated by ESO, AUI/NRAO and NAOJ.





\appendix

\bsp	
\label{lastpage}
\end{document}